\documentclass[12pt]{article}
\usepackage{graphicx}
\begin{document}


\begin{center}
{\Large \bf One anlytic form for four branches of the ABCD matrix}

\vspace{7mm}

S. Ba{\c s}kal \footnote{electronic address:
baskal@newton.physics.metu.edu.tr}\\
Department of Physics,
Middle East Technical University,
06531 Ankara, Turkey

\vspace{7mm}

Y. S. Kim\footnote{electronic address: yskim@umd.edu}\\
Center for Fundamental Physics, University of Maryland,\\
College Park, Maryland 20742

\end{center}

\vspace{6mm}

\begin{abstract}
 It is not always possible to diagonalize the optical $ABCD$
 matrix, but it can be brought into one of the four Wigner
 matrices by a similarity transformation.  It is shown that
 the four Wigner matrices can be combined into one matrix
 with four branches.  This result is illustrated in terms of
 optical activities, laser cavities, and multilayer optics.
\end{abstract}

\newpage

\section{Introduction}\label{intro}
The two-by-two $ABCD$ matrix plays a central role in optical
sciences. The four elements of this matrix are all real, and its
determinant is one.  Thus, it has three independent parameters.
Yet, it has a rich mathematical content which could lead
interesting results in physics.

It is generally assumed that this matrix can be diagonalized by
a rotation, but this is not the case as shown in our previous
paper~\cite{bk09josa}.  We have shown there that this matrix can
be brought to an equi-diagonal matrix by a rotation, and then by
a squeeze to one of the following four Wigner matrices.
\begin{equation}\label{wmat11}
\pmatrix{\cos\theta &  -\sin\theta \cr \sin\theta & \cos\theta},
                                                        \quad
\pmatrix{\cosh\lambda &  \sinh\lambda \cr
               \sinh\lambda & \cosh\lambda}, \quad
\pmatrix{1 & -\gamma \cr 0 & 1}, \quad
\pmatrix{1 & 0 \cr \gamma & 1}.
\end{equation}
This squeeze portion of the similarity transformation is not yet
widely known.  Thus the similarity transformation which brings the
$ABCD$ matrix to one of the Wigner matrices is a rotation followed
by a squeeze.  Even though the two triangular matrices in
Eq.(\ref{wmat11}) can be similarity-transformed from each other, it
is convenient to work with the four branches of the $ABCD$ matrix.

The purpose of this paper is to reduce these four matrices into one
analytic matrix with four different branches.  First of all, each
of the matrices in Eq.(\ref{wmat11}) is generated by
\begin{equation}\label{gen01}
\frac{1}{2} \pmatrix{0 & -i \cr i & 0}, \qquad
\frac{1}{2} \pmatrix{0 & i \cr i & 0}, \qquad
\frac{1}{2} \pmatrix{0 & -i \cr 0 & 0}, \qquad
\frac{1}{2} \pmatrix{0 & 0 \cr i & 0} ,
\end{equation}
respectively.  The last two matrices can be obtained from a linear
combination of the first two, with two independent coefficients.
We can then study the general property of the $ABCD$ matrix by
exponentiating the linear combination of the two matrices
\begin{equation}\label{gen02}
\pmatrix{0 & -i \cr i & 0} \quad \mbox{and} \quad
\pmatrix{0 & i \cr i & 0} ,
\end{equation}
and making Taylor expansions.

One of the present authors noted this aspect of the $ABCD$ matrix
while studying optical activities~\cite{kim09jmo}.  He then
concluded that the asymmetric optical activity can lead to the
study of the fundamental space-time symmetries of elementary
particles~\cite{wig39,knp86}.

In this paper, we study the resulting exponential form more
systematically.  We first exponentiate the linear combination
of these two independent matrices.  While the exponent is a fully
analytic function, the Taylor expansion of the exponential form
leads to complications, leading to four separate branches.

Again in this paper, we use the same optical activity to study
the origin of this branching property.  We note then that the
exponential form is convenient for repeated applications of the
$ABCD$ matrix, such as periodic systems including laser cavities
and multi-layer optics.

In Sec.~\ref{branch}, we discuss how the $ABCD$ matrix can be
written as an exponential function of one analytic matrix, with
four branches.  In Sec.~\ref{acti}, we use optical activities
to study the physics of the mathematics of Sec.~\ref{branch}.
Section~\ref{periodic} is devoted to application of this methods
to periodic systems.  Laser cavities and multilayer optics
are discussed in detail.

\section{Exponential Form and Branches}\label{branch}
Let us start with the $ABCD$ matrix as a rotated equi-diagonal
$abcd$ matrix:
\begin{equation}\label{abcd11}
[ABCD] = R(\alpha) [abcd] R(-\alpha) ,
\end{equation}
where $R(\alpha)$ is a rotation matrix
\begin{equation}\label{rot11}
R(\alpha) = \pmatrix{\cos(\alpha/2) & -\sin(\alpha/2) \cr
            \sin(\alpha/2) & \cos(\alpha/2)} ,
\end{equation}
and $[abcd]$ is an equi-diagonal matrix
\begin{equation}
[abcd] = \pmatrix{a & b \cr c & d} ,
\end{equation}
with
\begin{eqnarray}\label{alpha}
&{}& \tan\alpha = \frac{D - A}{B + C}, \nonumber\\[1ex]
&{}& a = d = \frac{A + B}{2} , \nonumber\\[1ex]
&{}& b = \frac{(B - C) + \sqrt{(A - D)^2 + (B + C)^2}}{2} ,
                                           \nonumber\\[1ex]
&{}& c = \frac{(C - B) + \sqrt{(A - D)^2 + (B + C)^2}}{2} .
\end{eqnarray}

Since the determinant of the $ABCD$ matrix is assumed to be one,
this matrix has three independent parameters.  One of those
parameters is the angle $\alpha$.  Thus, the $abcd$ matrix has
two independent parameters.  Since the two diagonal elements
of the $abcd$ matrix are the same, it can be exponentiated as
\begin{equation}\label{exp01}
[abcd] = \exp{\left\{r M(\theta)\right\}} ,
\end{equation}
with
\begin{equation}\label{m11}
M(\theta) =  \pmatrix{0 & -\cos\theta +
 \sin\theta \cr \cos\theta + \sin\theta & 0} .
\end{equation}
Here the two independent parameters are $r$ and $\theta$.
Thus, we are led to study in detail this $M(\theta)$ matrix
which can also be written as
\begin{equation}\label{gen03}
 M(\theta) = \pmatrix{0 & -1\cr 1 & 0} \cos\theta +
              \pmatrix{0 & 1\cr 1 & 0} \sin\theta .
\end{equation}
Other than the factor of $(i/2)$, this expression becomes the four
generators given in Eq.(\ref{gen01}) when $\theta = 0, \pi/2,
\pi/4, -\pi/4 $ respectively.

\begin{figure}[thb]
\centerline{\includegraphics[scale=0.5]{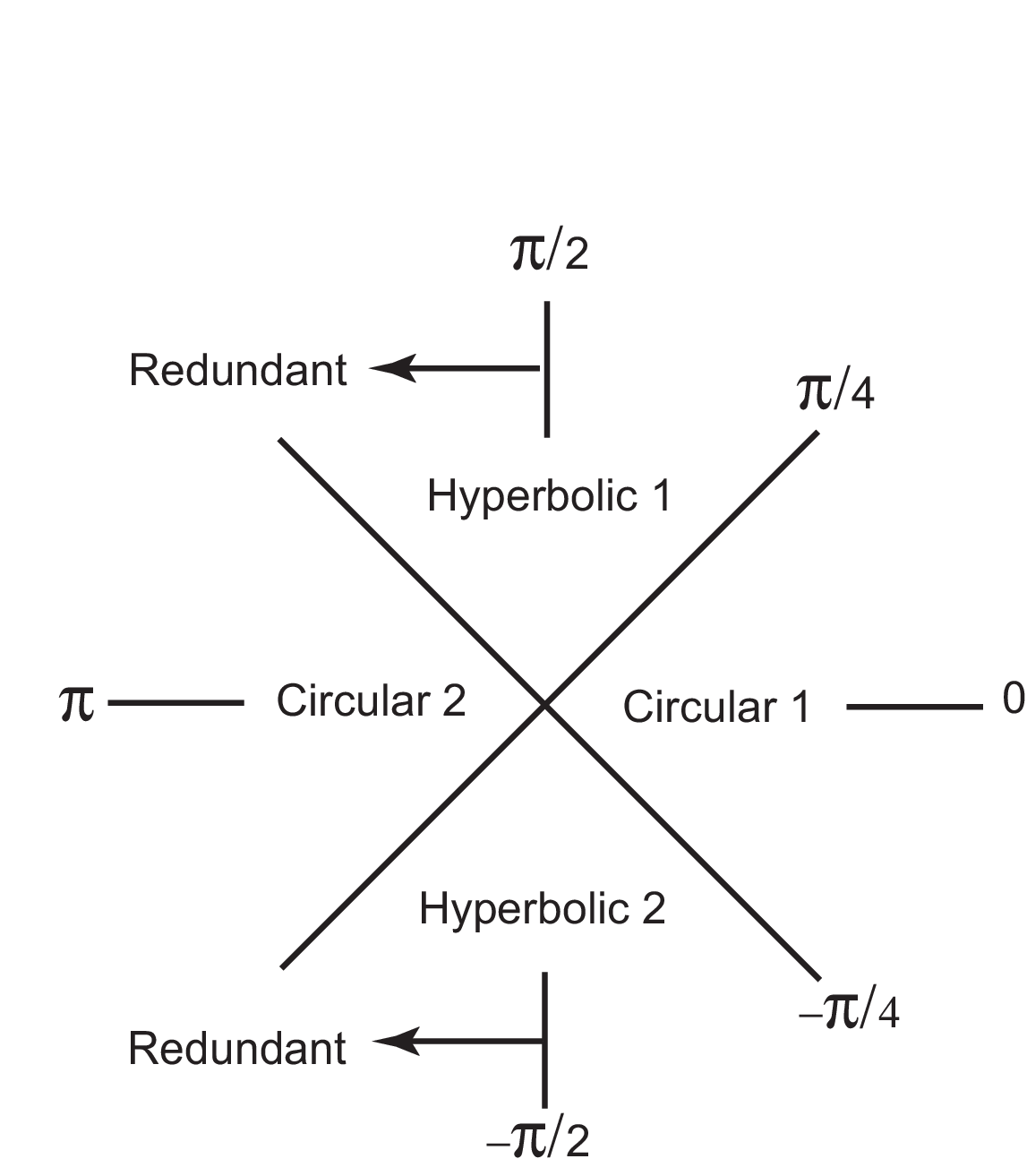}}
\vspace{6mm}
\caption{Forms of the $ABCD$ matrix depending on the angle
$\theta.$}\label{conic}
\end{figure}


In this way, we can combine the four Wigner matrices into an
exponential function of one analytic matrix.  The problem is how
to compute the exponential form of Eq.(\ref{exp01}).  Its Taylor
expansion is
\begin{equation}\label{taylor11}
[abcd] = \sum_{n} \frac{r^n}{n!} [M(\theta)]^n .
\end{equation}
This is an infinite series except at $\theta = \pm \pi/4$.  If
$\theta = 45^o$, the $M$ matrix becomes
\begin{equation}
M = \pmatrix{0 & 0 \cr \sqrt{2} & 0} .
\end{equation}
Since $M^2 = 0$, the series truncates.  The $abcd$ matrix becomes
\begin{equation}\label{trian11}
[abcd] = \pmatrix{1 & 0 \cr r\sqrt{2} & 1} .
\end{equation}
Likewise, when $\theta = -\pi/4$,
\begin{equation}\label{trian22}
[abcd] = \pmatrix{1 & -r\sqrt{2}  \cr 0 & 1} .
\end{equation}
This aspect is illustrated in Fig.~\ref{conic}.  The Taylor series
truncates at $\theta = \pm \pi/4$.  In the circular regions 1 and
2, $(\sin\theta)^2$ is smaller than $(\cos\theta)^2$.  In the
hyperbolic regions 1 and 2, $(\cos\theta)^2$ is smaller than
$(\sin\theta)^2$.

Also in Fig.~\ref{conic}, the symmetry of trigonometry  tells us
the region $\pi/2 < \theta < 3\pi/2 $ is redundant if we allow
both positive and negative values of $r$ in Eq.(\ref{exp01}).
In this way, we restrict $\cos\theta$ to positive values.  In
circular region 1, $\sin\theta$ can be both positive or negative.

In the region $(\cos\theta)^2 > (\sin\theta)^2,$ and $ |\theta| < \pi/4$,
if $\sin\theta$ is positive, we can write the $M$ matrix as
\begin{equation}
M(\theta) = \sqrt{\cos(2\theta)}
\pmatrix{0 & - \exp{(-\eta)} \cr \exp{(\eta)} & 0} .
\end{equation}
with
\begin{equation}\label{eta11}
\exp(-\eta) = \sqrt{\frac{\cos\theta - \sin\theta}
                     {\cos\theta + \sin\theta}} ,
\end{equation}
where $\eta$ is positive.
This formula is valid also when $\sin\theta$ is negative, but
$\eta$ is also negative.
We now write $M(\theta)$ as
\begin{equation}
M(\theta) = \sqrt{\cos(2\theta)}
\pmatrix{e^{-\eta/2} & 0 \cr 0 & e^{\eta/2}}
\pmatrix{0 & - 1 \cr 1 & 0}
\pmatrix{e^{\eta/2} & 0 \cr 0 & e^{-\eta/2}} .
\end{equation}
Then
\begin{equation}
 \left(M(\theta)\right)^n = \left(\cos(2\theta)\right)^{n/2}
\pmatrix{e^{-\eta/2} & 0 \cr 0 & e^{\eta/2}}
\pmatrix{0 & - 1 \cr 1 & 0}^n
\pmatrix{e^{\eta/2} & 0 \cr 0 & e^{-\eta/2}} .
\end{equation}
Thus
\begin{equation}
[abcd] = \pmatrix{e^{-\eta/2} & 0 \cr 0 & e^{\eta/2}}
\pmatrix{\cos\phi & -\sin\phi \cr \sin\phi  & \cos\phi}
\pmatrix{e^{\eta/2} & 0 \cr 0 & e^{-\eta/2}} ,
\end{equation}
which is
\begin{equation} \label{mat11}
[abcd] = \pmatrix{\cos\phi & -e^{-\eta}\sin\phi \cr
                  e^{\eta}\sin\phi  & \cos\phi}
\end{equation}
with
\begin{equation}\label{phi11}
\phi = r\sqrt{\cos(2\theta)} .
\end{equation}
In terms of the four parameters of the $ABCD$ matrix,
\begin{eqnarray}\label{phi22}
&{}& \cos\phi = \frac{A + B}{2} ,      \nonumber\\[1ex]
&{}& e^{-2\eta} = \frac{-b}{c} =
           \frac{C - B - \sqrt{(B + C)^2 + (A - D)^2}}
                {C - B + \sqrt{(B + C)^2 + (A - D)^2}} .
\end{eqnarray}
where $\cos\phi$ is smaller than one, and $b$ is negative.

If $(\sin\theta)^2 >(\cos\theta)^2$,  or
$ \pi/4 < |\theta| < \pi/2 $,  we have to consider two separate
regions in Fig.~\ref{conic}, where $\cos\theta$ is positive, while
$\sin\theta$ can take both positive and negative signs.
$\cos(2\theta)$ is negative.  The $M$ matrix should becomes
\begin{equation}\label{m22}
M(\theta) = \sqrt{-\cos(2\theta)}
\pmatrix{0 &  \exp{(-\eta)} \cr \exp{(\eta)} & 0} .
\end{equation}
with
\begin{equation}\label{eta22}
\exp(-\eta) =\sqrt{\frac{\sin\theta - \cos\theta}
              {\sin\theta + \cos\theta}} ,
\end{equation}
for both positive and negative values $\sin\theta$, but $\eta$ is
positive and is negative respectively.

Then the Taylor expansion leads to
\begin{equation}\label{mat22}
[abcd] = \pmatrix{\cosh\chi & e^{-\eta}\sinh\chi \cr
                  e^{\eta}\sinh\chi  & \cosh\chi} ,
\end{equation}
with
\begin{equation}\label{chi11}
\chi = r \sqrt{-\cos(2\theta)} .
\end{equation}
In terms of the parameters of the original $ABCD$ matrix,
\begin{eqnarray}
&{}& \cosh\chi = \frac{A + B}{2} ,      \nonumber\\[1ex]
&{}& e^{-2\eta} = \frac{b}{c} = \frac{B - C + \sqrt{(B + C)^2 + (A - D)^2}}
                {C - B + \sqrt{(B + C)^2 + (A - D)^2}} .
\end{eqnarray}
Here $\cosh\chi$ is greater than one, and both $b$ and $c$ are positive.

We can now go back to Eq.(\ref{eta11}) and Eq.(\ref{eta22}), and write
$\tan\theta$ in terms of $\eta$. Then $\tan\theta$ becomes
\begin{equation}
\tan\theta = \frac{b + c}{c - b} =
           \frac{\sqrt{(A - D)^2 + (B + C)^2}}{C  - B} ,
\end{equation}
for all values of $\theta$ between $-\pi/2$ and $\pi/2$.
The parameter $r$ is
\begin{eqnarray}
r = \left[\frac{b^2 + c^2}{-2bc}\right]^{1/2}\phi, \nonumber \\[1ex]
r = \left[\frac{b^2 + c^2}{2bc}\right]^{1/2} \chi ,
\end{eqnarray}
for $(\sin\theta)^2 < (\cos\theta)^2$ and
$(\sin\theta)^2 > (\cos\theta)^2$ respectively, with
\begin{equation}
\left[\frac{b^2 + c^2} {2bc}\right]^{1/2}
             = \left[\frac{2\left(B^2 + C^2\right) + (A - D)^2}
{4BC + (A - D)^2} \right]^{1/2}.
\end{equation}

Let us now look at how the transition of the $abcd$ from
Eq.~(\ref{mat11}) to Eq.~(\ref{mat22}).  This is a puzzling question
because the matrix $M(\theta)$ remains analytic in the neighborhood
of $\theta = \pi/4 $ (see Fig.~\ref{conic}. In order to tackle this
problem, we write $M(\theta)$ of Eq.(\ref{m11}) as
\begin{equation}
M(\theta) = (\cos\theta)
       \pmatrix{0  &  -(1 - \tan\theta) \cr 1 + \tan\theta & 0} .
\end{equation}
In the neighborhood of $\theta = \pi/4$, we can set $\cos\theta
= 1/\sqrt{2}$ and  $(1 + \tan\theta) = 2$, and
\begin{equation}
M(\theta) = \frac{1}{\sqrt2}
         \pmatrix{0  &  -(1 - \tan\theta) \cr  2 & 0} .
\end{equation}
Then up to $r^2$, the Taylor leads to
\begin{equation}
[abcd] =  = \pmatrix{1 - r^2(1 - \tan\theta)/2
    &  -r(1 - \tan\theta)/\sqrt{2}
        \cr r\sqrt{2}  &  1 - r^2(1 - \tan\theta)/2} .
\end{equation}
If $\theta$ is smaller than $\pi/4$, the diagonal elements of this
matrix are smaller than $1$, like $\cos\phi$ in Eq.(\ref{mat11}).
If $\theta$ becomes greater than $\pi/4$, the diagonal element
becomes greater than $1$ like $\cosh\chi$ in Eq.(\ref{mat22}).
If $\tan\theta = 1$, the result becomes that of
Eq.(\ref{trian11}).

We can give a similar reasoning for the neighborhood of
$\tan\theta = -1$.  The Taylor expansion leads to
\begin{equation}
[abcd] =
 \pmatrix{1 - r^2(1 + \tan\theta)/2 & -r\sqrt{2} \cr
 r(1 + \tan\theta)/\sqrt{2} & 1 - r^2(1 + \tan\theta)/2} .
\end{equation}
leading to Eq.(\ref{trian22}) for $\theta= -\pi/4.$

The exponential form given in Eq.(\ref{exp01}) is very convenient
when we study periodic systems where the $ABCD$ matrix is applied
repeatedly.  We shall return to this problem in
Sec.~\ref{periodic}.

\section{Optical Activities}\label{acti}
In his recent paper~\cite{kim09jmo}, one of the present authors used
the two-by-two matrix formulation of optical activities applicable
to the transverse electric field of an optical wave.  The direction
of the electric component rotates as the optical wave propagates.
In the real world, the medium causes also an attenuation of the
transverse components.  This does not interfere with the rotational
character.  However, there is a problem if the dissipation
coefficients are different for two perpendicular directions.

Let us start from a circularly polarized light wave which can be
decomposed into the right-polarized and left polarized components.
If they have different indexes of refraction, we can write the light
wave as
\begin{equation}
\pmatrix{E_{x} \cr E_{y}} =
\frac{1}{2}\pmatrix{ 1 \cr i}
 \exp{\left\{i \left(k_1 z - \omega t \right)\right\}}
+ \frac{1}{2}\pmatrix{ 1 \cr -i}
\exp{\left\{i \left(k_2 z - \omega t \right)\right\}}
\end{equation}
This two terms can be combined into
\begin{equation}\label{ray11}
\pmatrix{E_{x} \cr E_{y}} =
 \pmatrix{ \cos(\gamma z) \cr \sin(\gamma z)}
           \exp{\left\{i \left(k z - \omega t \right)\right\}},
\end{equation}
with
\begin{equation}
k = \frac{1}{2}\left(k_1 + k_2\right) , \qquad
                    \gamma = \frac{k_1 -k_2}{2} .
\end{equation}
If we start with a polarized light wave taking the form
\begin{equation}
\pmatrix{E_{x} \cr E_{y}} =
\pmatrix{A \exp{\left\{i(kz - \omega t)\right\}} \cr 0} ,
\end{equation}
the optical activity is carried out by the rotation matrix
\begin{equation}\label{rot22}
 R(\gamma z) = \pmatrix{\cos(\gamma z) & -\sin(\gamma z)  \cr
 \sin(\gamma z) & \cos(\gamma z)} .
\end{equation}

The optical ray is expected to be attenuated due to absorption
by the medium.  The attenuation coefficient in one transverse
direction could be different from the coefficient along the
other direction.  Thus, if the rate of attenuation along the
$x$ direction is different from that along $y$ axis, this
asymmetric attenuation can be described by
\begin{equation}\label{atten}
\pmatrix{\exp{\left(-\mu_{1}z \right)} & 0 \cr
   0 & \exp{\left(-\mu_{2}z \right)}}
 = e^{-\lambda z} \pmatrix{\exp{(\mu z)} & 0 \cr
                       0 & \exp{(-\mu z)}} ,
\end{equation}
with
\begin{equation}
\lambda = \frac{\mu_{2} + \mu_{1}}{2} , \qquad
      \mu = \frac{\mu_{2} - \mu_{1}}{2} .
\end{equation}
The exponential factor $\exp{(-\lambda z)}$ is for the overall
attenuation, and the matrix
\begin{equation} \label{sq01}
 \pmatrix{\exp{(\mu z)} & 0 \cr 0 & \exp{(-\mu z)}}
\end{equation}
performs a squeeze transformation.  This matrix expands the $x$
component of the polarization, while contracting the $y$ component.
We shall call this the squeeze along the $x$ direction.

The squeeze does not have to be along the $x$ and $y$directions
For convenience,  let us rotate the squeeze axis by $45^o$.  Then
the squeeze matrix becomes
\begin{equation}\label{sq02}
S(\mu z) = \pmatrix{\cosh(\mu z) & \sinh(\mu z) \cr
 \sinh(\mu z) & \cosh(\mu z) } .
\end{equation}

If this squeeze is followed by the rotation of Eq.(\ref{rot22}),
the net effect is
\begin{equation}
e^{-\lambda z} \pmatrix{\cos(\gamma z) & -\sin(\gamma z) \cr
 \sin(\gamma z) & \cos(\gamma z)}
\pmatrix{\cosh(\mu z) & \sinh(\mu z) \cr
 \sinh(\mu z) & \cosh(\mu z) } ,
\end{equation}
where $z$ is in a macroscopic scale, perhaps measured in
centimeters.  However, this is not an accurate description
of the optical process.

This happens in a microscopic scale of $z/N$, and becomes
accumulated into the macroscopic scale of $z$ after the $N$
repetitions, where $N$ is a very large number.  We are thus
led to the transformation matrix of the form
\begin{equation}\label{trans}
Z(\gamma,\mu,z)= \left[e^{-\lambda z/N}S(\mu z/N)
                  R(\gamma z/N)\right]^N .
\end{equation}
In the limit of large $N$, this quantity becomes
\begin{equation}
e^{-\lambda z} \left[\pmatrix{1 & \mu z/N \cr \mu z/N & 1}
\pmatrix{1 & - \gamma z/N \cr \gamma z/N & 1}\right]^N .
\end{equation}
Since $\gamma z/N$ and $\mu z/N$ are very small,
\begin{equation}\label{z11}
 Z(\gamma,\mu,z)= e^{-\lambda z} \left[\pmatrix{1 & 0 \cr 0 & 1}
  + \pmatrix{0 & - \gamma + \mu \cr
 \gamma + \mu & 0}\frac{z}{N} \right]^N .
\end{equation}
For large $N$, we can write this matrix as~\cite{kim09jmo}
\begin{equation}\label{expo22}
Z(\gamma,\mu, z) = e^{-\lambda z} \exp{\left\{kz M(\theta) \right\}} ,
\end{equation}
where the $M$ matrix is
\begin{equation}\label{m33}
 M(\theta) = \pmatrix{0 & -\cos\theta + \sin\theta
   \cr \cos\theta + \sin\theta & 0} ,
\end{equation}
 with
\begin{eqnarray}\label{gm11}
&{}& k ={\sqrt{\gamma^2 + \mu^2}}, \nonumber \\[1ex]
&{}& \cos\theta = \frac{\gamma}{\sqrt{\gamma^2 + \mu^2}}, \nonumber \\[1ex]
&{}& \sin\theta = \frac{\mu}{\sqrt{\gamma^2 + \mu^2}}.
\end{eqnarray}
We note here that the $M(\theta)$ matrix of Eq.(\ref{m33}) is the
same as that of Eq.(\ref{m11}) which determines the branch property
of the $ABCD$ matrix.  At this point, it is more convenient to work
with $kM(\theta)$.
\begin{equation}
kM(\theta) = \pmatrix{0 & -\gamma + \mu \cr \gamma + \mu & 0} .
\end{equation}

If $\gamma > \mu$, the $rM$ matrix can then be written as
\begin{equation}
 kM = \sqrt{\gamma^2 - \mu^2}\pmatrix{0 & -e^{-\eta}
               \cr e^{\eta} & 0},
\end{equation}
where $\eta$ of Eq.(\ref{eta11}) becomes
\begin{equation}
 e^{-2\eta} = \sqrt{\frac{\gamma - \mu}{\gamma + \mu}} =
 \sqrt{\frac{\cos\theta - \sin\theta}{\cos\theta + \sin\theta}} .
\end{equation}
Thus, the exponential function in Eq.(\ref{expo22}) can be evaluated
according to the procedure defined in Sec.~\ref{branch}.  This
expression is the same as that of Eq.(\ref{eta11}).

The exponential form $\exp{(kzM)}$ in of Eq.(\ref{expo22})
becomes
\begin{equation}
 \pmatrix{e^{-\eta/2} & 0 \cr 0 & e^{\eta/2}}
 \pmatrix{\cos(\gamma' z) & -\sin(\gamma' z) \cr
                    \sin(\gamma'z)  & \cos(\gamma' z)}
 \pmatrix{e^{\eta/2} & 0 \cr 0 &  e^{-\eta/2}} ,
\end{equation}
where
\begin{equation}\label{gammak}
\gamma' = \sqrt{\gamma^2 - \mu^2} .
\end{equation}
The transformation matrix of Eq.(\ref{expo22}) takes the form
\begin{equation}
Z(\gamma,\mu,z) = e^{-\lambda z}
      \pmatrix{\cos(\gamma' z) & - e^{-\eta} \sin(\gamma' z) \cr
      e^{\eta} \sin(\gamma' z)  & \cos(\gamma' z)} ,
\end{equation}

If $\mu > \gamma$, the $rM$ matrix becomes
\begin{equation}
 kM = \sqrt{\mu^2 - \gamma^2}\pmatrix{0 & e^{-\eta} \cr e^{\eta} & 0},
\end{equation}
where $\eta$ of Eq.(\ref{eta22}) takes the form
\begin{equation}
 e^{-2\eta} = \sqrt{\frac{\mu - \gamma}{\mu + \gamma}} =
    \sqrt{\frac{\sin\theta - \cos\theta}
     {\cos\theta + \sin\theta}},
\end{equation}
and $Z$ becomes
\begin{equation}
Z(\gamma,\mu,z) = e^{-\lambda z}
      \pmatrix{\cosh(\mu'z) &  e^{-\eta} \sinh(\mu'z) \cr
          e^{\eta} \sinh(\mu' z)  & \cosh(\mu' z)} ,
\end{equation}
where
\begin{equation}\label{muk}
\mu' = \sqrt{\mu^2 - \gamma^2}.
\end{equation}

In this section, we discussed a system of optical activities with
asymmetric dissipation as a physical illustration of the
mathematical procedure discussed in Sec.~\ref{branch}.  We have
already seen that $M(\theta)$ of Eq.(\ref{m33}) has the same
form as that of Eq.(\ref{m11}), and that angle $\theta$ can be
defined in terms of the parameters $\gamma$ and $\mu$, as shown
in Eq.(\ref{gm11}).  The parameter $\eta$ can also be defined
in terms of $\gamma$ and $\mu$, and its expression is the
same as the one given in terms of the angle $\theta$.

As for the branches, we note that both $\gamma$ and $\mu$ can be
negative and positive.  Thus, the angle $\theta$ can cover the
entire range from zero to $2\pi$.  We can write $\gamma'$ and
$\mu'$ as
\begin{equation}
\gamma' = k\sqrt{\cos^2\theta - \sin^2\theta}, \qquad
\mu' = k \sqrt{\sin^2\theta - \cos^2\theta} .
\end{equation}
Since $\cos^2\theta - \sin^2\theta = \cos(2\theta),$
$\gamma' z$ and $\mu' z$ correspond to $\phi$ and $\chi$ of
Eq.(\ref{phi11}) and Eq.(\ref{chi11}) respectively.

If we start with $\mu = 0$, it is simply a rotation of the
transverse component of the electric field and the overall
attenuation factor is $\exp{(-\lambda z)}$.  The rate of this
rotation decreased as $\mu$ increases, and the rotation stops at
$\gamma = \mu$.  For $\mu > \gamma$, there are no rotations.  It
would be very interesting to test these effects experimentally.

We should not forget the fact that the equi-diagonal
$[abcd]$ matrix is a rotated $ABCD$ matrix.  The rotation
matrix is given in Eq.(\ref{rot11}).  This rotation changes
the optical ray of Eq.(\ref{ray11}) to
\begin{equation}\label{ray22}
\pmatrix{E_{x} \cr E_{y}} =
 \pmatrix{ \cos(\gamma z + \alpha/2) \cr
                              \sin(\gamma z + \alpha/2)}
           \exp{\left\{i \left(k z - \omega t \right)\right\}}.
\end{equation}
This is also an observable effect.

We have seen in this section that the asymmetric optical activity
can serve as an analog computer for the mathematical procedure
given in Sec.~\ref{branch} which is in fact an alternative to the
diagonalization of the $ABCD$ matrix.

\section{Periodic Systems in Optics}\label{periodic}

 Let us summarize what we can do about the $ABCD$ matrix.
\begin{itemize}

\item[1.] We should first rootate to an equi-diagonal matrix [abcd].

\item[2.] If the diagonal elements of this equi-diagonal matrix are
   smaller than one, it can be written as
    \begin{equation}
      \pmatrix{\cos\phi & -e^{-\eta} \sin\phi \cr
                  e^{\eta} \sin\phi & \cos\phi } ,
    \end{equation}
    with $\exp{(-\eta)} = -b/c$, which can also be written in
    terms of the elements of the original $ABCD$ matrix, as shown
    in Eq.~(\ref{phi11}).

\item[3.] If the diagonal elements of the equi-diagonal matrix are
    greater than than one, the matrix can be written as
    \begin{equation}
      \pmatrix{\cosh\chi & e^{-\eta} \sinh\chi \cr
                  e^{\eta} \sinh\chi & \cosh\chi } ,
    \end{equation}
    with $\exp{(-\eta)} = b/c$, which takes the form of
    Eq.(\ref{chi11}) in terms of the elements of the $ABCD$ Matrix.

\item[4.]  If one of the off-diagonal elements vanish, the diagonal
     elements have to be one.

\item[5.]  It is possible to combine all these cases into one
     exponential function of one analytic matrix.  It can be
     written as
 \begin{equation}
 [abcd] = \exp{ \left\{r\pmatrix{0  & -\cos\theta + \sin\theta \cr
         \cos\theta + \sin\theta & 0} \right\}} .
 \end{equation}

\item[6.] When $\theta = \pm\pi/4,$  the Taylor series truncates,
   and
   \begin{equation}
    [abcd] =  \pmatrix{1 &  -r(1 \mp 1)/\sqrt{2} \cr
       r(1 \pm 1)/\sqrt{2} & 1 } .
   \end{equation}

\end{itemize}

\subsection{Laser Cavities}\label{cav}

\begin{figure}[thb]
\centerline{\includegraphics[scale=0.4]{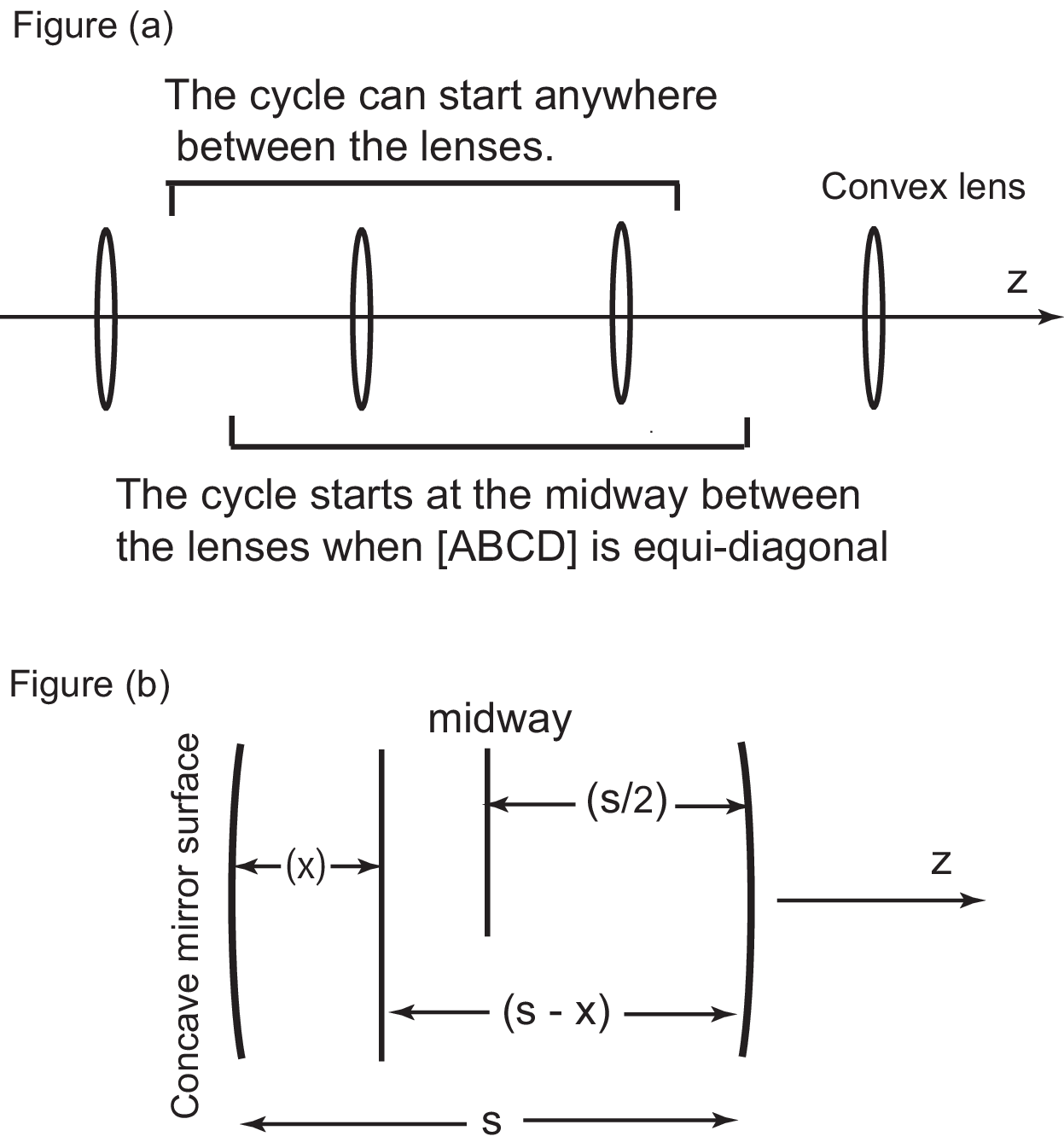}}
\vspace{5mm}
\caption{Optical rays in a laser cavity.
(a) Multiple cycles in a laser cavity are equivalent to the beam
going through multiple lenses, for which one cavity cycle corresponds
to the propagation of light through a sub-system of two lenses.  The
$ABCD$ matrix becomes equi-diagonal when the cycle begins at the midway
between the lenses.  (b) A laser cavity consisting of two concave
mirrors with separation $s$.}\label{cavity22}

\end{figure}

A laser cavity consists of two concave mirrors separated by distance
$s$ as illustrated in Fig.~\ref{cavity22}.  The mirror matrix takes
the form
\begin{equation}\label{lens01}
\pmatrix{1 & 0 \cr -2/R & 1} ,
\end{equation}
where $R$ is the radius of the concave mirror.  The separation matrix
is
\begin{equation}\label{lens02}
\pmatrix{1 & s \cr 0 & 1} .
\end{equation}
If we start the cycle from one of the two mirrors
one complete cycle consists of
\begin{equation}\label{lens03}
\pmatrix{1 & 0 \cr -2/R & 1} \pmatrix{1 & s \cr 0 & 1}
\pmatrix{1 & 0 \cr -2/R & 1} \pmatrix{1 & s \cr 0 & 1} .
\end{equation}
If we start the beam at the position $x$ from the mirror, then one
complete cycle becomes
\begin{eqnarray}\label{lens04}
&{}&\pmatrix{1 & x \cr 0 & 1} \pmatrix{1 & 0 \cr -2/R & 1}
 \pmatrix{1 & s -x \cr 0 & 1}     \nonumber \\[1ex]
&{}& \hspace{20mm}\times \pmatrix{1 & x \cr 0 & 1}
\pmatrix{1 & 0 \cr -2/R & 1}\pmatrix{1 & s - x \cr 0 & 1} .
\end{eqnarray}
This cycle consist of two identical half cycles.

Thus, we shall use the half-cycle matrix as our starting point.
Then the half-cycle $ABCD$ matrix becomes
\begin{equation}\label{lens05}
[ABCD] = \pmatrix{1 & x \cr 0 & 1} \pmatrix{1 & 0 \cr -2/R & 1}
  \pmatrix{1 & s - x \cr 0 & 1},
\end{equation}
It is now possible to replace replace $R$ and $x$ and by $R/s$
and $x/s$ respectively and set $s = 1$~\cite{bk09josa}.  Then
\begin{equation}\label{lens07}
[ABCD] = \pmatrix{ 1 - 2xf &  1 - 2xf(1 - x) \cr -2f  & 1 - 2f(1 - x)} ,
\end{equation}
where $f = s/R$, and is expected to be a small number because the
mirror radius $R$ is much larger than the separation of the mirrors.

It can be brought to an equi-diagonal form by a rotation as given in
Eq.(\ref{abcd11}).  According to Eq.(\ref{alpha}), the rotation angle
is
\begin{equation}
      \tan\alpha = \frac{2f(2x - 1)}{1 - 2f\left(1 + x - x^{2}\right)} .
\end{equation}
This angle is zero when $x = 1/2$.  In this case, the laser cycle
starts at the midway between the lenses~\cite{bk09josa}.  Then the
$ABCD$ matrix becomes
\begin{equation}\label{lens09}
[abcd] = \pmatrix{1 - f &  1 - f/2 \cr - 2f & 1 - f} ,
\end{equation}
This matrix can then be written as
\begin{equation}
[abcd] = \pmatrix{\cos\phi &  e^{\eta}\sin\phi \cr
      -e^{-\eta}\sin\phi & \cos\phi} .
\end{equation}
with
\begin{equation}\label{phieta}
\cos\phi = 1 - f, \qquad
e^{2\eta} = \frac{2 - f}{4f} .
\end{equation}
This is the result we obtained in our earlier paper on laser
cavities~\cite{bk02}, where the cycle starts from the midway
between the lenses.  The signs of $\phi$ and $\eta$ are opposite
to those given in Eq.(\ref{mat11}), but this is purely for
convenience.  There are no fundamental problems.

We can now  write this expression in an exponential form
\begin{equation}
[abcd] = \exp{\left\{r\pmatrix{0 & \cos\theta + \sin\theta \cr
      -\cos\theta + \sin\theta & 0 }\right\}},
\end{equation}
with
\begin{eqnarray}
&{}& \tan\theta = \frac{2 - 5f}{2 + 3f} , \nonumber \\[1ex]
&{}&  r = \left[\frac{13f -17f^2}{8 - 17f^2}\right]^{1/2} \phi ,
\end{eqnarray}
where $\phi$ is given in Eq.(\ref{phieta}).
Since the radius of the mirror is much larger than the mirror
separation, $f$ is a small number, and $\tan\theta$ is close to
one and $r$ is a small number.

The $N$-cycle laser consists of $2N$ half-cycles, and its $abcd$
matrix is
\begin{equation}
[abcd]^{2N} = \exp{\left\{2Nr\pmatrix{0 & \cos\theta + \sin\theta \cr
       -\sin\theta + \cos\theta & 0}\right\}} .
\end{equation}

This section is a straight-forward application of the procedure given
in Sec.~\ref{branch}.  We know that $\sin(r\theta)^{2N}$ is not
$\sin(2Nr\theta)$, but the exponential form gives us the convenience of
$[\exp{(ir\theta)}]^{2N} = \exp{(i2Nr\theta)}.$  We have given a
two-by-two matrix formulation of this convenience applicable to the
$ABCD$ matrix.

\subsection{Multilayer Optics}\label{multi}

From the physical concept of Wigner's little group whose transformations
leave the four-momentum of a given particle invariant~\cite{wig39,knp86},
it has been established in the literature that~\cite{gk03}
\begin{equation}\label{lg11}
S(\eta) W S(-\eta) = R(\xi) B(-2\lambda) R(\xi) ,
\end{equation}
where $W$ is one of the four Wigner matrices given in Eq.(\ref{wmat11}),
$R(\xi)$ is the rotation matrix of the form of Eq.(\ref{rot11}),
and
\begin{eqnarray}\label{sqz11}
&{}& S(\eta) = \pmatrix{e^{\eta/2}  & 0 \cr 0 & e^{-\eta/2}},
\nonumber\\[1ex]
&{}& B(\lambda) = \pmatrix{\cosh(\lambda/2) & \sinh(\lambda/2) \cr
           \sinh(\lambda/2) & \cosh(\lambda/2)} ,
\end{eqnarray}
and the continuous parameters $\xi$ and $\lambda$ take care of the four
different Wigner parameters.  These parameters can be written in terms
of $\eta$ and the parameter of the Wigner matrix, as shown in
Ref.~\cite{bk09josa,gk03}.

Since the left side of Eq.(\ref{lg11}) can be written as an exponential
form, we can write
\begin{equation}
R(\xi) B(-2\lambda)R(\xi) = \exp{\left\{ r \pmatrix{0 &
 -\cos\theta + \sin\theta \cr \cos\theta + \sin\theta  & 0}\right\}} ,
\end{equation}
where $r$ and $\theta$ are also continuous variables.

\begin{figure}[thb]
\centerline{\includegraphics[scale=0.5]{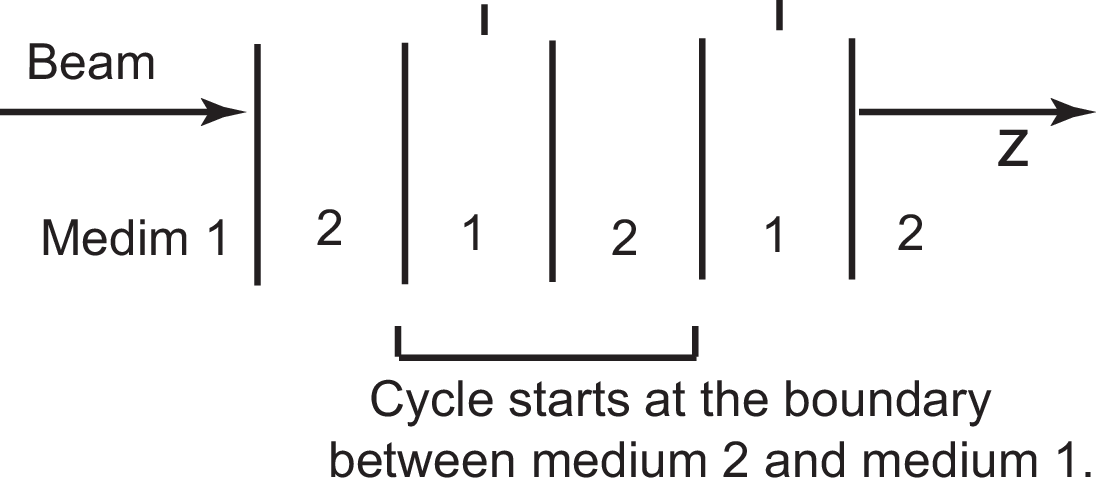}}
\caption{Multilayer consisting of two different refractive indices.
One complete cycle starts at the boundary between medium 2 and
medium 1.}\label{mlayer22}
\end{figure}

With this mathematical preparation, let us study multilayer optics.
In this branch of optics, we have to consider the $ABCD$ matrix
applicable to two beams moving in opposite directions, one which is
the incident beam and the other is the reflected beam~\cite{azzam77}.
We can represent them as
a two component column matrix
\begin{equation}
\pmatrix{ E_{+}e^{ikz} \cr E_{-} e^{-ikz}} ,
\end{equation}
where the upper and lower components correspond to the incoming
and reflected beams respectively.
For a given frequency, the wave number depends on the index of the
refraction.  Thus, if the beam travels along the distance $d$, the
column matrix should be multiplied by the two-by-two
matrix~\cite{azzam77}
\begin{equation}\label{ps11}
P(\beta_{j})=\pmatrix{e^{i\beta_{j}/2} & 0 \cr 0 & e^{-i\beta_{j}/2}} ,
\end{equation}
where $\beta_{j}/2 = k_{j}d $ and $j$ is denoting each different medium.
If the beam propagates along the first medium and meets the boundary
at the second medium, it will be partially reflected and partially
transmitted.  The boundary matrix is~\cite{azzam77}
\begin{equation}\label{bn1}
B(\nu)=\pmatrix{\cosh(\nu/2)  &  \sinh(\nu/2)  \cr
               \sinh(\nu/2)  &  \cosh(\nu/2) } ,
\end{equation}
with
\begin{equation}
\cosh(\nu/2) = 1/t_{12}, \qquad \sinh(\nu/2) = r_{12}/t_{12} ,
\end{equation}
where $t_{12}$ and $r_{12}$ are the transmission and reflection
coefficients respectively, and they satisfy
$\left(r_{12}^2 + t_{12}^2\right) = 1.$
The boundary matrix for the second to first medium is the
inverse of the above matrix.
Therefore, one complete cycle, starting from the second medium,
consists of
\begin{equation}
  B(\nu)P(\beta_{1})B(-\nu)P(\beta_{2}),
\end{equation}
as illustrated in Fig.~\ref{mlayer22}.  This complex-valued
matrix can be cast into a real matrix by a similarity
transformation with the transformation matrix
\begin{equation} \label{cmatrix}
C=\frac{1}{\sqrt{2}}
  \pmatrix{e^{i\pi/4} &  e^{i\pi/4}  \cr
       - e^{-i\pi/4} &  e^{-i\pi/4}} ,
\end{equation}
This transforms the boundary matrix $B(\nu)$ of Eq.(\ref{bn1}) to a
squeeze matrix $S(\nu)$ of Eq.(\ref{sqz11}), and the phase shift
matrices $P(\beta_{j})$ of Eq.(\ref{ps11}) to rotation matrices
$R(\beta_{j})$ of the form given in Eq.(\ref{rot11}).  We are thus
led to consider the $ABCD$ matrix of the form
\begin{equation}\label{abcd44}
[ABCD] = S(\nu)R(\beta_{1})S(-\nu)R(\beta_{2}).
\end{equation}

If $W$ in Eq.(\ref{lg11}) is a rotation matrix, we can write
\begin{equation}\label{lg22}
 S(\nu)R(\beta_{1})S(-\nu) = R(\xi_{1})B(-2\lambda)R(\xi_{1})
\end{equation}
where
\begin{eqnarray}
&{}&\cosh \lambda = (\cosh\nu) \sqrt{1-\cos^{2} (\beta_{1}/2)
\tanh^{2}\nu}, \\
&{}& \cos\xi_{1}= \frac{\cos(\beta_{1})}
{(\cosh\nu)\sqrt{1-\cos^{2}(\beta_{1}/2)\tanh^{2}\nu}}.
\end{eqnarray}
The $ABCD$ matrix can then be simplified to
\begin{equation}\label{bd22}
[ABCD] = R(\xi_{1})B(-2\lambda)R(\xi_{2})
\end{equation}
with
\begin{equation}
\xi_{2} = \xi_{1} + \beta_{2}
\end{equation}

It is now possible to write the above form as
\begin{equation}
[ABCD] = R(\alpha)[R(\xi)B(-2\lambda)R(\xi)]R(-\alpha) ,
\end{equation}
with
$$ \xi = \frac{1}{2} \left(\xi_1 + \xi_2\right), \qquad
     \alpha = \frac{1}{2} \left(\xi_1 - \xi_2\right).
$$
The role of the rotation matrix $R(\alpha)$ matrix is clearly
defined in Sec.~\ref{branch}.  Thus $R(\xi)B(-2\lambda)R(\xi)$
is the equi-diagonal matrix, and
\begin{equation}
R(\xi)B(-2\lambda)R(\xi) = \pmatrix{\cosh\lambda\cos\xi
                 & -(\sin\xi\cosh\lambda + \sinh\lambda) \cr
                \sin\xi~\cosh\lambda - \sinh\lambda &
                \cosh\lambda~\cos\xi}  .
\end{equation}
Thus, if $(\cosh\lambda \cos\xi)$ is smaller than one, we can
write this matrix as
\begin{equation}
\pmatrix{\cos\phi & -e^{\eta}\sin\phi \cr
   e^{-\eta}\sin\phi  & \cos\phi} ,
\end{equation}
with
\begin{equation}
\cos\phi = (\cosh\lambda)\cos\xi,  \qquad
e^{2\eta} = \frac{(\cosh\lambda)\sin\xi + \sinh\lambda}
                {(\cosh\lambda)\sin\xi - \sinh\lambda} .
\end{equation}
Thus, if $(\cosh\lambda \cos\xi)$ is greater than one, we should
write the equi-diagonal matrix as
\begin{equation}
\pmatrix{\cosh\chi &  -e^{\eta}\sinh\chi \cr
   -e^{-\eta}\sinh\chi  & \cosh\chi} ,
\end{equation}
with
\begin{equation}
\cosh\chi = (\cosh\lambda)\cos\xi,  \qquad
e^{2\eta} = \frac{\sinh\lambda + (\cosh\lambda)\sin\xi}
                {\sinh\lambda - (\cosh\lambda)\sin\xi} .
\end{equation}

We are now interested in the exponential form
\begin{equation}\label{expo66}
R(\xi)B(-2\lambda)R(\xi)
= \exp{\left\{r\pmatrix{0 & - (\cos\theta + \sin\theta)  \cr
\cos\theta - \sin\theta & 0}\right\}} .
\end{equation}
with
\begin{equation}
\tan\theta = \frac{\tanh\lambda}{\sin\xi} .
\end{equation}
As for the $r$ parameter,
\begin{equation}
 r = \left[\frac{\sin^2\xi + \tanh^2\lambda}
   {\sin^2\xi - \tanh^2\lambda} \right]^{1/2} \phi , \quad
 r = \left[\frac{\sin^2\xi + \tanh^2\lambda}
   {\tanh^2\lambda - \sin^2\xi} \right]^{1/2} \chi ,
\end{equation}
for $(\cosh\lambda \cos\xi) < 1$, and for
$(\cosh\lambda \cos\xi) > 1$, respectively.

In this section, we started with two media with two different
indexes of refraction, corresponding to two rotation matrices
$R\left(\beta_{1}\right)$ and $R\left(\beta_{2}\right)$ given
in Eq.(\ref{abcd44}) respectively.  However, the combined effect
in not necessarily a rotation matrix.  It can be analytically
continued to the hyperbolic branch through the exponent of the
$ABCD$ matrix.

When $(\cosh\lambda \cos\xi)^2 = 1$, one of the off-diagonal
elements in Eq.(\ref{expo66}) vanishes, and this case was
repeatedly discussed in the literature~\cite{bk09josa,gk03},
also in the present paper.

\section*{Concluding Remarks}

In this paper, we noted first that the two-by-two $ABCD$ matrix
can be represented as a similarity transformation of one of the
four matrices which we choose to call the Wigner matrices.  We
then combined these Wigner matrices into one exponential form of
an analytic matrix.

While $\cos\phi$ and $\cosh\chi$ correspond to a circle and a
hyperbola respectively, the lines in Fig.~\ref{conic} correspond
parabolas in the four-dimensional representation of the Lorentz
group~\cite{wig39,kiwi90jm}.  Ancient Greeks used a circular cone
to combine these curves into one.  This is the reason why we call
them conic sections.  It is gratifying to note that the optical
devices we discussed in Secs.~\ref{acti} and~\ref{periodic} can
play the role of a conic section.  Instead of three-dimensional
cone, we used a two-dimensional plane in Fig.~\ref{conic}.

We have seen in this paper that Taylor expansion of this analytic
form results in four branches.  How does this happen? Let us go
to the Taylor expansion of Eq.(\ref{taylor11}).  This infinite
series truncates at $(\sin\theta)^2 = (\cos\theta)^2$ or along
the two lines in Fig.~\ref{conic}.  We are not familiar with
mathematical singularities resulting from the truncation of the
infinite Taylor series.  This appears to be an interesting problem
in mathematics, but it is beyond the scope of this paper.


\begin{thebibliography}{99}

\bibitem{bk09josa}
S. Ba{\c s}kal and Y. S. Kim, J. Opt. Soc. Am.  A {\bf  26},
 3049-2054 (2009).

\bibitem{kim09jmo}
Y. S. Kim, J. Mod. Opt. {\bf 57}, 17-22 (2010).

\bibitem{wig39}
E. Wigner,  Ann. Math. {\bf 40}, 149-204 (1939).

\bibitem{knp86}
Y. S. Kim and M. E. Noz, {\em Theory and Applications of the
 Poincar\'e Group} (Reidel, Dordrecht, 1986).

\bibitem{bk02}
S. Ba{\c s}kal and Y. S. Kim, Phys. Rev. E {\bf 66}, 026604-026609
(2002).

\bibitem{gk03}
E. Georgieva and Y. S. Kim, Phys. Rev. E {\bf 68}, 026606: 1-6 (2003).

\bibitem{azzam77}
R. M. A. Azzam and N. M. Bashara, {\em Ellipsometry and Polarized
 Light} (Elsevier, Amsterdam, 1977).

\bibitem{kiwi90jm}
Y. S. Kim and E. P. Wigner, J. Math. Phys. {\bf 31}, 55-60 (1990).

\end{thebibliography}
\end{document}